\documentstyle[aps,twocolumn]{revtex}

\input{epsf}

\begin{document}
\author{D. Blume, and Chris H. Greene}
\title{
Quantum corrections to the ground state energy of a trapped 
Bose-Einstein condensate: 
A diffusion Monte Carlo calculation}
\date{\today }
\address{Department of Physics and JILA, University of Colorado, 
Boulder, CO 80309-0440, USA}
\maketitle

\begin{abstract}
The diffusion Monte Carlo method is applied
to describe a trapped atomic
Bose-Einstein condensate at zero temperature, 
fully quantum mechanically and nonperturbatively.
For low densities, $n(0)a^3 \le 2 \cdot 10^{-3}$ [$n(0)$: peak density, 
$a$: $s$-wave scattering length], 
our calculations confirm that the exact ground state
energy for a sum of two-body interactions depends on 
only the atomic physics parameter $a$, 
and no other details of the
two-body model potential.
Corrections to the mean-field Gross-Pitaevskii 
energy range from being essentially negligible to 
about 20~\% for  
$N=2-50$ particles in the trap with positive $s$-wave scattering length
$a=100-10000$~a.u..
Our numerical calculations confirm that
inclusion of an additional effective potential term
in the mean-field equation, which accounts for
quantum fluctuations 
[see e.g. E. Braaten and A. Nieto, Phys. Rev. B {\bf{56}}, 14745 (1997)], 
leads to a greatly improved
description of trapped Bose gases.
\end{abstract}

\draft

\section{Introduction}
\label{introduction} 
Since the first achievement of Bose-Einstein condensation 
in trapped atomic vapors in 1995~\cite{ande95}, these systems 
have received increased attention from both experimental
and theoretical efforts.
Theoretical studies of these inhomogeneous gases are primarily
based on the mean-field Gross-Pitaevskii (GP) equation
for the condensate
wave function~\cite{dalf98}.  This equation can also be viewed
as a variant of the Hartree-Fock (HF)
approximation~\cite{esry97}.
It is thus fundamentally important for our understanding of this 
many-boson system to ascertain the validity of the mean-field 
description and the importance of particle correlations.
In principle, these questions can be investigated through the use of 
a complete many-body basis instead of making a
single particle approximation.  Some studies have
in fact considered
the $T=0$ condensate 
at the level of the random phase approximation (RPA), with a few 
going still further to treat the particle-particle correlations at
the level of configuration interaction 
(CI)~\cite{esry97,esry99,holz98,holz99a}.
For many particles, however,
a full CI calculation exceeds current computational capabilities.  
Consequently we adopt the
diffusion Monte Carlo (DMC) method for the calculations reported in 
this paper.

In the mean-field description, the particle interactions enter
solely through an effective mean-field potential, which 
is proportional to the $s$-wave scattering 
length 
$a$~\cite{dalf98,bogo47,pita61,gros61,gros63,huan57,lee57,lee57a,huan63,fett71}.
In contrast, a full description treats a many-body
potential surface and allows nonseparability of the many-body
wave function.
The first key question of our study is therefore:
Can the $s$-wave scattering length approximation properly
describe an inhomogeneous Bose gas, or does the actual
form of the two-body
interaction potential necessarily come into play?
To test the validity
of the shape independent atom-atom potential 
(frequently expressed as a $\delta$-function
potential with $s$-wave scattering length $a$~\cite{ferm34}), 
we solve the $N$-body Schr\"odinger equation for an
inhomogeneous Bose gas for 
various different two-body
potentials that generate identical $a$
using the DMC method for densities $n(0)a^3 \le 2 \cdot 10^{-3}$.
We find that different potentials produce indistinguishable 
total ground state energies $E$, 
indicating that the lowest many-body Schr\"odinger energy eigenvalue 
is independent 
of the shape of the two-body potential.
This result is in agreement with predictions based
on an expansion in $\sqrt{n(0)a^3}$ for low densities~\cite{braa97,timm97}.

However, we do observe differences between the
accurate many-body ground state energy and
that obtained by solving the GP 
equation.
A second motivation of our study is therefore 
to assess the validity of the mean-field description 
as a function of $a$ and $N$, and as a function of the 
peak density $n(0)a^3$.
Furthermore, we test the validity of a modified GP equation 
that accounts for quantum fluctuations~\cite{braa97,ande99}.
In addition to the ground state energy, we also show 
that the condensate density can be significantly affected by
correlations.
These calculations support our conclusion that the modified 
GP equation [see Eq.~(\ref{GP_mod})] greatly improves upon 
the commonly used mean-field treatment.

Theoretical studies of correlation effects
in trapped condensates
have already been 
proposed~\cite{holz98,holz99a,braa97,timm97,ande99,krau96,grut97,fabr99,dubo00,zalu00,zieg97,hutc98,morg00}.
These approaches include approximate zero temperature
variational Monte Carlo studies~\cite{dubo00},
essentially exact finite temperature path integral Monte
Carlo studies~\cite{holz98,holz99a,krau96,grut97} 
as well as correlated basis function approaches~\cite{fabr99}, 
and perturbative schemes~\cite{braa97,timm97,ande99,morg00}.
Here, we use essentially exact
diffusion Monte Carlo
techniques to directly calculate correlation effects at
zero temperature. 

Section~\ref{system} introduces the many-body Schr\"odinger equation 
of a trapped Bose gas and the numerical treatment applied to solve this
equation.
Results and their interpretation are presented in Section~\ref{results}.
Section~\ref{summary} summarizes this paper.

\section{System and numerical techniques}
\label{system} 
The many-body Schr\"odinger equation for a condensate 
of $N$ mass $m$ bosons in a 
spherical trap, centered at the origin
with trapping frequency $\omega_{ho}$, is given by
\begin{equation}
	\left(-\frac{\hbar^2}{2m} \sum_i^N \nabla_i^2 +
	\sum_{i<j}^N V(r_{ij}) +
	\sum_i^N \frac{1}{2} m \omega_{ho}^2 \vec{r}_i^2\right) \psi=E\psi.
	\label{many-body}
\end{equation}
Here $ V(r_{ij})=V( |\vec{r}_i-\vec{r}_j|)$ denotes the two-body
interaction potential, and $\vec{r}_i = (x_i,y_i,z_i)$ is the Cartesian
position vector of atom $i$ relative to the trap center.
$m$ is taken to be $m(^{87}$Rb), and
$\omega_{ho}$ to be $ 2 \pi \times 77.78$~Hz.
In the following we
use harmonic oscillator units
for energies ($\hbar \omega_{ho}$)
and length
($a_{ho}=[\hbar/(m \omega_{ho})]^{1/2}$).
The energy per particle for the
non-interacting case, $V(r)=0$, is then $E/N =1.5$~$\hbar \omega_{ho}$,
and, for example,
the $^{87}$Rb $s$-wave triplet scattering length ($a \approx 100$~a.u.), 
$a=0.00433$~$a_{ho}$ at that frequency.

The DMC method~\cite{hamm94} (and references therein),
here implemented with importance sampling and a descendant weighting
scheme, 
derives a solution to the  
many-body ground state Schr\"odinger equation, Eq.~(\ref{many-body}),
for a given model potential
surface. This solution is essentially exact, apart from statistical 
uncertainties. The resulting 
ground state energy shall be denoted by $E_{DMC}$ in the 
following.
Ideally, one would solve Eq.~(\ref{many-body}) 
for the best known $^{87}$Rb-$^{87}$Rb two-body 
potential $V(r)$ (e.g. Ref.~\cite{krau90}).
However, the large number of bound states in this potential
makes its use in a many-body calculation
extremely challenging, if not infeasible (see below).
Instead, we have carried out tests using three different 
two-body model potentials A-C to test how the energetics
depend
on the actual form of the two-body potential:
A) a purely repulsive 
hard core potential with hard core radius $a$,
$V(r) = \infty$ for $r<a$ and $V(r)=0$ for $r \ge a$; 
B) another purely repulsive potential with parameters
$d > 0$ and $r_0$, namely $V(r)=d \cosh^{-2}(r/r_0)$; and
C) a sum of an attractive and a repulsive
Gaussian, leading to a potential that exhibits
a minimum with negative energy.
The potential parameters are adjusted such that 
{\it i}) $a=0.00433$~$a_{ho}$
(potentials A-C),
and {\it ii}) $a=0.433$~$a_{ho}$ (potentials A and B).

Potential C requires a few more remarks.
In contrast to potentials A and B, which are purely repulsive, potential C is
repulsive at short range and 
attractive at long range.
The usage of this potential in a DMC
calculation could therefore potentially lead to 
a lowest energy state 
that describes a cluster rather than a condensed state.
However, the two-body potential is such that it does not support
a two-body bound state; 
nevertheless, many-body bound states might exist.
If this two-body potential supports a many-body bound state, then our
simulation should converge to a cluster state
rather than to the metastable condensed state at infinite 
imaginary simulation time. At finite time, however, we find empirically
for positive $a$
that our DMC simulation samples the condensed and not the cluster  state.

To test our DMC code we carry out a separate 
exact solution to 
Eq.~(\ref{many-body}) for two particles, $N=2$,
by separating out the center of mass motion and solving the 
one-dimensional radial Schr\"odinger equation using 
standard numerical techniques. 
For potential A we find, for example, $E=3.00345$~$\hbar \omega_{ho}$ for 
$a=0.00433$~$a_{ho}$, and 
$E=3.3827$~$\hbar \omega_{ho}$ for $a=0.433$~$a_{ho}$ 
in excellent agreement with our DMC results,
$E_{DMC}=3.00346(1)$~$\hbar \omega_{ho}$ and 
$E_{DMC}=3.3831(7)$~$\hbar \omega_{ho}$, respectively.
(The statistical error of the DMC energies is given in parenthesis.)
Since
the DMC algorithm generalizes straightforwardly
as a function of $N$,
the above agreement is a convincing 
test of our code.
Furthermore, we carefully checked the independence of
our results on the exact shape
of the trial wave function and on the time step
used for the propagation in imaginary time. 
The next section presents results from our DMC calculation and compares 
them with GP theory.

\section{Results and interpretation}
\label{results} 
Columns 2-4 of Table~\ref{tab1}
show the DMC energies $E_{DMC}$ for $N=3-20$
for the three two-body potentials A, B and C with
{\it i}) $a=0.00433$~$a_{ho}$ (top) and 
{\it ii}) $a=0.433$~$a_{ho}$ (bottom).
For the parameter range considered in Table~\ref{tab1}, 
which corresponds to $n(0)a^3 \le 2 \cdot 10^{-3}$ (see below),
the ground state energy $E_{DMC}$
of the inhomogeneous gas for different $V(r)$ with identical
two-body scattering length $a$ is independent of the shape of $V(r)$; and
thus $E_{DMC}$ depends only on $a$ to within our statistical uncertainties.
For the homogeneous Bose gas, 
Giorgini~{\em et al.}~\cite{gior99}
find a small dependence of the 
ground state energy on the shape of the two-body potential
in the intermediate density regime, 
$na^3 \approx 10^{-3}$ ($n$: number density of homogeneous gas);
however, their energy depends only on $a$ in the low density regime.
Recent extensive studies of trapped two- and three-particle condensates
as a function of the scattering length $a$ and the trap frequency 
$\omega_{ho}$
reveal dependences of the ground and excited state energies
on the exact shape of the two-body potential in the high
density limit. However, these studies will be presented
elsewhere.

To ascertain the validity of the widely used mean-field 
approach,
we now compare our many-body DMC energies with those 
obtained by solving the GP equation,
\begin{eqnarray}
	\left[
	\frac{-\hbar^2}{2m} \nabla^2 +
	\frac{1}{2} m\omega_{ho}^2 \vec{r}^2 +
	\frac{4 \pi \hbar^2 (N-1)a}{m} |\Phi_{GP}(\vec{r})|^2 \right ]  
	\nonumber \\
	\times \Phi_{GP}(\vec{r}) = \epsilon_{GP} \Phi_{GP}(\vec{r}).
	\label{GP}
\end{eqnarray}
Here, $\Phi_{GP}(\vec{r})$ denotes the
ground state orbital (normalized to 1), and 
$\epsilon_{GP}$ the orbital energy; the total
energy
$E_{GP}$ can be obtained through evaluation of the energy functional 
$E_{GP}[\Phi_{GP}]$~\cite{dalf98}.
Notice that $\epsilon_{GP}$ can also be associated with the chemical potential
of the system.
Both $\epsilon_{GP}$ and $E_{GP}$ depend only on $(N-1)a$,
rather than on $a$ and $N$ separately.
The formulation of the GP equation in terms of
$(N-1)a$ rather than $Na$ follows from number conserving
Schr\"odinger quantum mechanics, using a
HF initial state~\cite{esry97}.
To compare our DMC energies $E_{DMC}$ with 
the GP energies $E_{GP}$, our many-body DMC
calculations reported  below
have been performed for 
potential A only, which is the
purely repulsive hard core potential.
Since $E_{DMC}$ is independent
of the shape of $V(r)$ (see Table~\ref{tab1}),
this justifies our restriction to only one two-body potential.
Note, however, that our calculations reported in the following
treat densities $n(0)a^3$ as large as $\approx 1$.
In this high density regime we expect 
the many-body energy to depend on the 
detailed shape of the two-body potential.

\begin{figure}[tbp]
\centerline{\epsfysize=4.in\epsfbox{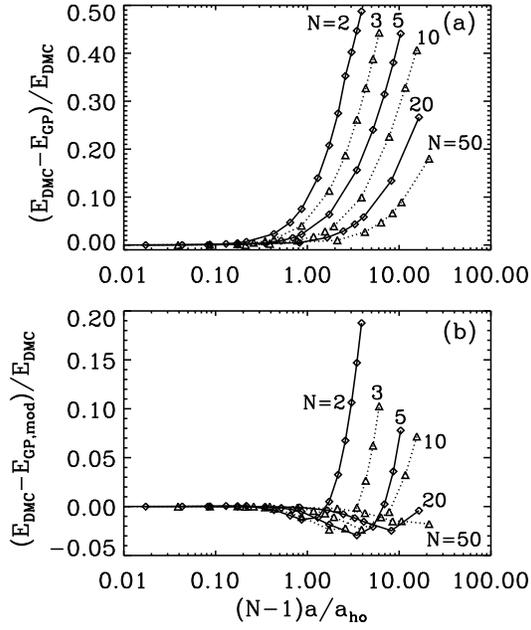}}
\caption{Energy difference 
$E_{DMC}-E_{GP}$ normalized by $E_{DMC}$ (a) and 
$E_{DMC}-E_{GP,mod}$ normalized by $E_{DMC}$ (b), respectively,
as a function of the GP interaction parameter $(N-1)a/a_{ho}$.
Diamonds and triangles show data points resulting from solving 
Eqs.~(\ref{many-body})-(\ref{GP_mod}), and solid and 
dotted lines connect 
data points for identical $N$ to guide the eye, $N=2-50$.
Note the logarithmic $(N-1)a/a_{ho}$ scale.
} 
\label{fig1}
\end{figure}

Figure~\ref{fig1}(a) compares the ground state energies 
$E_{GP}$ and $E_{DMC}$
for $N=2-50$ as a function of 
the interaction parameter $(N-1)a/a_{ho}$ [Eq.~(\ref{GP})].
Note the logarithmic scale for $(N-1)a/a_{ho}$;
diamonds and triangles 
show our calculated data points, while solid and dotted lines connect those
data points with identical $N$ but different $a$.
The full quantum energies $E_{DMC}$ are always equal to or larger than the 
mean-field
energies $E_{GP}$, with
$(E_{DMC}-E_{GP})/E_{DMC}$ attaining $20$ to $50$~\% for 
the largest scattering lengths $a$ considered here,
$a=3.9,3.0,2.6,1.7,0.87,0.43$~$a_{ho}$ for $N=2,3,5,10,20,50$.
Figure~\ref{fig1}(a) shows
the separate
dependences of 
$E_{DMC}$ on $a$ and $N$; e.g.,
for $(N-1)a=10$~$a_{ho}$,
$(E_{DMC}-E_{GP})/E_{DMC} \approx 0.4$ for $N=5$, 
but instead $\approx 0.07$ for $N=50$.
Column~5 of Table~\ref{tab1} summarizes $E_{GP}$
for a few selected $N$ and $a$ parameters.
In conclusion, 
the GP equation, Eq.~(\ref{GP}), describes the condensate accurately
for small $a$. For larger $a$, however, significant 
departures occur from the DMC results.
Later in this section we discuss these departures further.

We refer to the energy difference 
$E_{DMC}-E_{GP}$ as a ``correlation energy'' $E_{corr}$.
Note that the definition of $E_{corr}$ introduced
here differs from that often used in quantum chemistry, where
$E_{corr}$ is defined as the energy
difference between the CI
energy $E_{CI}$ and the HF energy $E_{HF}$
for the same two-body potential $V(r)$.
In our treatment
$E_{DMC}$ results from solving the
many-body Schr\"odinger equation,
Eq.~(\ref{many-body}), for the exact wave function
(= complete basis)
with a {\it{shape dependent}} two-body potential $V(r)$, whereas $E_{GP}$
results from the mean-field equation, Eq.~(\ref{GP}), or equivalently,
from Eq.~(\ref{many-body}) for a {\it{shape independent}} $\delta$-function
potential using a HF initial state.
Thus,
$E_{corr}$ not only contains basis set effects, 
but also effects due to
the reduction of the shape dependent 
interaction potential to a shape independent potential
that generates an identical scattering length $a$.
Note that $E_{DMC} \ge E_{GP}$, and therefore $E_{corr} \ge 0$,
for the parameter range considered here.
The ``exact'' many-body treatment of the particle interaction 
(using a model two-body potential) leads evidently 
to an {\it{increase}} of the mean-field energy $E_{GP}$, which
is somewhat counterintuitive.
Recent calculations for potentials of type B 
with negative potential depth $d$ show, however, that
the many-body energy can be lower than the GP energy
for extremely tight traps, $\nu_{ho} \approx 10$~MHz 
(see, e.g., Fig.~3 of~\cite{esry99}). In the following we restrict 
ourselves to
the hard core potential A.

Next, we assess a modified GP equation, which has been discussed in 
the literature~\cite{braa97,timm97}.
Braaten and Nieto~\cite{braa97} included the effects of
quantum fluctuations for positive $a$ to generate 
corrections to the mean-field equation. 
Qualitatively, quantum fluctuations can be viewed as
representing virtual double excitations of
particles out of the condensate, which is also related to the 
ground state depletion.
Within the Thomas-Fermi (TF) approximation,
these ``quantum corrections'' produce an additional local
effective potential term in the GP equation that is linear in
the assumed small parameter 
$(n a^3)^{1/2}$,
\begin{eqnarray}
	\bigg[
	\frac{-\hbar^2}{2m} \nabla^2 +
	\frac{1}{2} m\omega_{ho}^2 \vec{r}^2 +
	\frac{4 \pi \hbar^2 (N-1)a}{m} |\Phi_{GP,mod}(\vec{r})|^2  \nonumber \\
	\times
	\left(1 + \frac{32}{3 \sqrt{\pi}} a^{\frac{3}{2}} (N-1)^{\frac{1}{2}} 
	\Phi_{GP,mod}(\vec{r})\right)
	 \bigg] \Phi_{GP,mod}(\vec{r}) = \nonumber \\
	\epsilon_{GP,mod} \Phi_{GP,mod}(\vec{r}).
	\label{GP_mod}
\end{eqnarray}
The total energy $E_{GP,mod}$ relevant to this modified 
GP equation can be obtained from the energy
functional $E_{GP,mod}[\Phi_{GP,mod}]$.
Equation~(\ref{GP_mod}) accounts for two-body physics through the 
scattering length $a$ solely;
three-body effects are not included.
In contrast to Eq.~(\ref{GP}), which only depends on the interaction parameter 
$(N-1)a$, Eq.~(\ref{GP_mod}) exhibits separate
dependences on $N-1$ and $a$.
Note that 
inclusion of second order terms in the TF expansion leads to additional
non-local terms in Eq.~(\ref{GP_mod}), which account for edge 
effects~\cite{braa97}.
Our estimates suggest that these non-local terms
are small,
and we have therefore neglected them.

The term accounting for quantum
fluctuations for the inhomogeneous gas, 
the second term in the second line of Eq.~(\ref{GP_mod}),
is identical to the Huang-Lee-Yang correction term for 
the homogeneous gas~\cite{huan57,lee57,lee57a}.
Huang, Lee and Yang showed in 1957 that the many-body Schr\"odinger
equation for a Bose gas interacting via hard sphere potentials can be 
equivalently formulated by replacing the hard sphere boundary conditions
by a pseudo-potential.
This pseudo-potential can then be treated in perturbation theory,
leading to a lowest order ground state energy per particle
$E_0/N \approx2 \pi\hbar^2  (a/m) n$.
To next order in $(n a^3)^{1/2}$,
the first order perturbation energy
can be evaluated 
by summing over momentum states with 
wave vector $\vec{k} \ne 0$~\cite{lee57a},
resulting in 
$E_0/N \approx 2 \pi \hbar^2 (a/m) n 
[1+\frac{128}{15 \sqrt{\pi}}(n a^3)^{1/2}]$
for the homogeneous gas. 
The summation over $\vec{k}$ states depends crucially on use of
the exact pseudo-potential expansion:
Use of $4 \pi \hbar^2 (a /m) \delta(\vec{r})$ rather
than $4 \pi \hbar^2 (a /m) \delta(\vec{r}) \frac{\partial}{\partial r}r$
leads to a divergent ground state energy expression.
The additional term arising in first order perturbation theory for the
homogeneous gas is identical
to the term in the inhomogeneous gas accounting for quantum fluctuations
(using quantum mean-field theory language)~\cite{braa97},
and directly shows the contribution of $k \ne 0$ (excited) states
to the ground state energy.
Braaten and Nieto~\cite{braa97} stress
that the most important excitations are to orbitals
in the energy range $E_{excitation} \approx 8 \pi \hbar^2 (a/m) n $.

The derivation of Eq.~(\ref{GP_mod}) is based on two assumptions~\cite{braa97}:
{\em i)} the expansion parameter $(n a^3)^{1/2}$
is small,
and
{\em ii)} $1/\sqrt{n a}$ is small compared to the condensate radius.
To assess the accuracy of these assumptions explicitly, Fig.~\ref{fig1}(b)
compares $E_{GP,mod}$ obtained from
Eq.~(\ref{GP_mod}) with our DMC energy $E_{DMC}$.
Consider the $N=2$ curve first.
$E_{DMC}-E_{GP,mod}$ is positive for small $a$, 
switches sign at $a \approx 0.2$~$a_{ho}$, reaches its
minimum for $a \approx 0.87$~$a_{ho}$, 
and becomes positive for $a > 1.7$~$a_{ho}$.
The behavior for larger $N$ is similar.
The maximal negative deviation 
for the parameter range shown in Fig.~\ref{fig1}(b)
is about $3$~\% for $N=5$
and $a \approx 0.87$~$a_{ho}$.
For $N=20$ and $N=50$, the agreement between $E_{DMC}$ and $E_{GP,mod}$ 
is better than 3~\% for $a$ as large as 
$0.87$~$a_{ho}$ ($\approx 20000$~a.u.) for $N=20$,
and $0.43$~$a_{ho}$ ($\approx 10000$~a.u.) for $N=50$, respectively.
Table~\ref{tab1} summarizes $E_{GP,mod}$ for a few selected 
values of $N$
and $a$.

The modified GP equation, Eq.~(\ref{GP_mod}),
overcorrects the correlation effects for some $a$ for all $N$ considered.
Overall,
the additional effective potential term in
Eq.~(\ref{GP_mod}) improves the GP treatment substantially for $N=2-50$.
As the size of the condensate increases, we expect the 
next higher order correction terms, which are non-local~\cite{braa97},
to become even smaller, and the local correction term considered 
in Eq.~(\ref{GP_mod}) to be the dominant one.
It should be noted, however, that 
Eq.~(\ref{GP_mod}) is derived in the large $N$ regime, where
the number of particles contributing to 
quantum fluctuations is small compared to those 
occupying the ``single particle ground state''~\cite{braa97}.
Thus
the excellent agreement between $E_{GP,mod}$ and $E_{DMC}$
in the small $N$ limit 
might be somewhat fortuitous.

\begin{figure}[tbp]
\centerline{\epsfysize=4.in\epsfbox{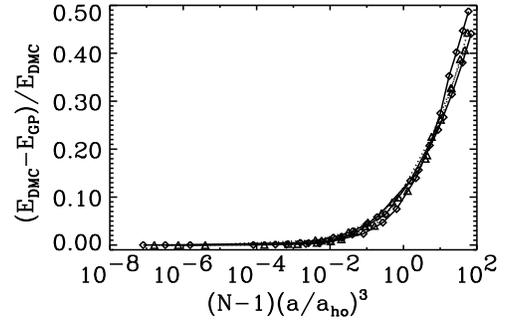}}
\caption{Energy difference 
$E_{DMC}-E_{GP}$ normalized by $E_{DMC}$, also shown
in Fig.~\protect\ref{fig1}(a), however, now   
as a function of $(N-1)(a/a_{ho})^3$ rather than
the interaction parameter $(N-1)a/a_{ho}$.
All data points, $N=2-50$, line up on a ``universal curve''.
Note the logarithmic $(N-1)(a/a_{ho})^3$ scale.
} 
\label{fig2}
\end{figure}

Figure~\ref{fig2} shows the same data as in Fig.~\ref{fig1}(a), however, now
as a function of $(N-1)(a/a_{ho})^3$ rather than the interaction parameter
$(N-1)a/a_{ho}$.
All data points, $N=2-50$, lie on a ``universal curve''.
As we shall discuss later, this universal behavior suggests that our
calculations for small $N$ and their
comparison with mean-field and quantum mean-field treatment 
may directly transfer to inhomogeneous gases
with larger $N$.

To summarize our ground state energy studies, 
note the different scales of the vertical axes in Figs.~\ref{fig1}(a)
and (b). A comparison of (a) and (b) 
shows that a significant 
improvement over GP theory is accomplished
through the inclusion of quantum
fluctuations [Eq.~(\ref{GP_mod})], as evidenced by the
better agreement between $E_{GP,mod}$ and $E_{DMC}$ than
between $E_{GP}$ and $E_{DMC}$.

\begin{figure}[tbp]
\centerline{\epsfysize=2.6in\epsfbox{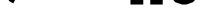}}
\caption{DMC density per particle $n_{DMC}(r)/N$ together with GP density
per particle
$n_{GP}(r)/N$ and the modified GP
density per particle $n_{GP,mod}(r)/N$ for $N=3$ (a) and $N=10$ (b) for
two different scattering lengths $a$.
Triangles show
the DMC density for $a=0.0433$~$a_{ho}$, 
diamonds that for $a=0.433$~$a_{ho}$.
Dashed lines show 
the GP and dotted lines
the modified GP densities per particle.
The statistical error of the DMC data is smaller
than the size of the symbols at large $r$, and about 
twice the size of the symbols near the trap center (small $r$).
Note $n_{GP}(r)/N$ and $n_{GP,mod}(r)/N$
are indistinguishable on the scale
shown for $N=3$ and $a=0.0433$~$a_{ho}$.
}
\label{density}
\end{figure}

Besides the ground state energy, we also calculate structural
properties using Eqs.~(\ref{many-body})-(\ref{GP_mod}).
The peak density $n(0)$, for example, determines
many properties such as the lifetime of a trapped condensate,
and its exact determination is therefore of great importance.
Figure~\ref{density} summarizes our results for
the radial density per particle $n(r)/N$ for $N=3$ (a) and $N=10$ (b)
for two different scattering lengths $a$, where 
$\int n(r) d^3\vec{r} =N$.
Our DMC density per particle
$n_{DMC}(r)/N$ [Eq.~(\ref{many-body})] is
shown as triangles for $a=0.0433$~$a_{ho}$  and as
diamonds for $a=0.433$~$a_{ho}$.
Note that these data points have statistical noise.
Dashed lines indicate the GP density per particle $n_{GP}(r)/N$ 
[Eq.~(\ref{GP})],
while dotted lines show the modified GP density per particle 
$n_{GP,mod}(r)/N$ 
[Eq.~(\ref{GP_mod})].
For the smaller scattering length, $a=0.0433$~$a_{ho}$, all
three densities per particle,
$n_{DMC}(r)/N$, $n_{GP}(r)/N$ and $n_{GP,mod}(r)/N$,
agree well. In fact, for $N=3$ the modified GP density and 
the GP density are indistinguishable.
For the larger scattering length,
$a=0.433$~$a_{ho}$, $n_{GP,mod}(r)/N$ coincides with the
DMC density $n_{DMC}(r)/N$, whereas the GP equation overestimates
the peak density by about $20$~\% ($N=3$), 
and $35$~\% ($N=10$), respectively.
For comparison, the GP treatment underestimates the energy 
for this scattering length by
only 4~\% (10~\%), whereas the modified GP
equation overestimates the energy by only about 2~\% (2~\%)
for $N=3$ ($N=10$).

\begin{figure}[tbp]
\centerline{\epsfysize=4.in\epsfbox{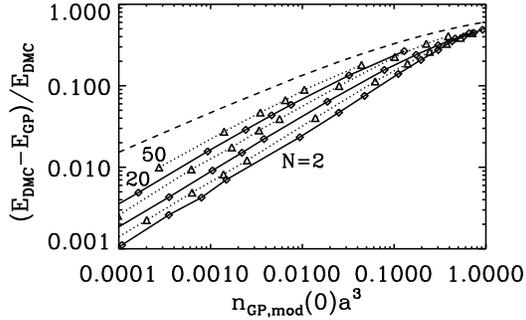}}
\caption{Energy difference 
$E_{DMC}-E_{GP}$ normalized by $E_{DMC}$, also shown
in Fig.~\protect\ref{fig1}(a) and \protect\ref{fig2}, however, now   
as a function of $n_{GP,mod}(0)a^3$ (solid and dotted lines)
for densities $10^{-4} \le n_{GP,mod}(0)a^3 \le 1$
together with the fractional correction to the GP energy
calculated within quantum mean-field theory using the TF approximation
(dashed line, see text).
Note the logarithmic scale of both axes.
} 
\label{fig4}
\end{figure}

The calculation of the DMC density $n_{DMC}(r)$,
and especially the peak density $n_{DMC}(0)$, 
is more time intensive than that of the DMC energy $E_{DMC}$.
To replot
our data from Fig.~\ref{fig2}
we therefore approximate $n_{DMC}(0)$
by $n_{GP,mod}(0)$, which is easy to calculate.
Figure~\ref{fig4} shows the fractional energy correction
$(E_{DMC}-E_{GP})/E_{DMC}$ on a log-log scale as a function of 
$n_{GP,mod}(0)a^3$ for $10^{-4} \le n_{GP,mod}(0)a^3 \le 1$
(solid and dotted lines).
In addition, the dashed line shows the fractional correction
to the GP energy predicted within quantum mean-field theory using the
TF (large $N$) limit,
$x/(1+x)$,
where $x= 7/8 \sqrt{\pi n(0)a^3}$~\cite{dalf98,braa97,timm97}.
The larger $N$, the smaller the difference between the 
fractional correction to the GP energy based on our DMC calculation
for small $N$
and the asymptotic large $N$ limit (dashed curve).
Figure~\ref{fig4} seems to suggest that the predicted behavior 
based on quantum mean-field theory within the TF 
limit~\cite{dalf98,braa97,timm97} is
indeed correct, and gives an accurate estimate of the error of the GP 
equation as a function of the density $n(0)a^3$ for large $N$.
Our calculations may therefore be viewed as a first explicit numerical
test of the application of quantum mean-field theory to 
trapped Bose gases.

\section{Summary}
\label{summary} 
This paper presents full quantum calculations  for an inhomogeneous
Bose gas that 
serve as a stringent test of  
mean-field theory.
Our calculations suggest that condensates are well described through
only the $s$-wave scattering length $a$.
Other characteristics of the two-body potential such as the 
effective range parameter seem to be neglegible for the 
$a$ and $N$ parameter range considered here. These findings 
for the inhomogeneous gas are in 
agreement with quantum mean-field predictions~\cite{braa97,ande99,timm97},
and recent DMC studies for the homogeneous Bose gas~\cite{gior99}.
Our studies also suggest that the first order correction term
[Eq.~(\ref{GP_mod})] leads to a great improvement upon the 
GP equation, [Eq.~(\ref{GP})], for large $a$.
For a spherical trap with, for example, $N=5000$ particles  
and $a= 0.0433$~$a_{ho}$ (1000 a.u.)
the GP and the modified GP energies differ noticeably, at the
$5.9$~\% level: $E_{GP}/N= 9.239$~$\hbar \omega_{ho}$, and 
$E_{GP,mod}/N = 9.786$~$\hbar \omega_{ho}$.
The densities for this condensate differ at the 15~\% level:
$n_{GP}(0)a^3= 1.9 \cdot 10^{-3}$ and $n_{GP,mod}(0)a^3= 1.7 \cdot 10^{-3}$.
We expect the excitation frequencies to change by an amount
comparable to that of the energy~\cite{pita98,braa99a}.
Very recently, stable condensates with large scattering
length $a=10000$~a.u., corresponding to a density
$n(0) a^3 = 10^{-2}$,
have been obtained experimentally in a non-spherical trap~\cite{corn00}.

The studies presented here are performed for a two-body potential
with positive
scattering length $a$,
which supports no two-body bound state. 
It will be interesting to extend these studies to 
two-body potentials with negative $a$.
Furthermore, the interplay between 
recombined molecular (i.e. ``snow flake'') and 
BEC-type states, completely neglected in this work, promises rich physics.

\section{Acknowledgements}
This work was supported partly by the National Science Foundation. 
D.~B. acknowledges support
through a DFG Postdoktorandenstipendium. Extensive discussions with 
John L. Bohn, E. Braaten and B. D. Esry have been greatly appreciated.

\begin{table}[tbp]
\begin{tabular}{cccccc}
 $N$ & potential A & potential B  & potential C & $E_{GP}$ &$E_{GP,mod}$ 
\\ \hline
 3 & 4.51036(2) & 4.51037(2) & 4.51035(4) & 4.51032 &4.51032 \\
 5 & 7.53443(4) & 7.53439(6) & 7.53441(10) & 7.53432 & 7.53434\\
10 & 15.1537(2) & 15.1539(2) & 15.1536(8)    & 15.1534 & 15.1535      \\
20 & 30.640(1) & 30.639(1) & 30.638(2) & 30.638 & 30.639\\ \hline
 3 & 5.553(3)   & 5.552(2) &  & 5.329 & 5.611 \\
 5 & 10.577(2)  & 10.574(4) &  & 9.901 &10.772\\
10 & 26.22(8) & 26.20(8) &  & 23.61 &26.84\\
20 & 66.9(4) & 66.9(1) &  & 57.9 & 68.5 \\ 
\end{tabular}
\caption{Ground state energies $E_{DMC}$, columns 2-4, 
in units of $\hbar \omega_{ho}$ for three different 
two-body potentials A-C for $N=3-20$ particles in
the trap; top for $a=0.00433$~$a_{ho}$ (potentials A-C), and bottom
for $a=0.433$~$a_{ho}$ (potentials A-B).
The statistical uncertainty is given in parenthesis.
For comparison columns 5 and 6 contain the GP energy $E_{GP}$ and
the modified GP energy $E_{GP,mod}$, respectively.}
\label{tab1}
\end{table}


\begin{thebibliography}{10}

\bibitem{ande95}
M.~H. Anderson, J.~R. Ensher, M.~R. Matthews, C.~E Wieman and E.~A. Cornell,
\newblock Science {\bf{269}}, 198 (1995).

\bibitem{dalf98}
F. Dalfovo, S. Giorgini, L.~P. Pitaevskii and S. Stringari,
\newblock Rev. Mod. Phys. {\bf{71}}, 463  (1999).

\bibitem{esry97}
B.~D. Esry,
\newblock Phys. Rev. A {\bf{55}},   1147 (1997).

\bibitem{esry99}
B.~D. Esry and C.~H. Greene,
\newblock Phys. Rev. A {\bf{60}},  1451 (1999).

\bibitem{holz98}
M. Holzmann, W. Krauth and M. Naraschewski,
\newblock Phys. Rev. A {\bf{59}},  2956 (1999).

\bibitem{holz99a}
M. Holzmann and Y. Castin,
\newblock Eur. Phys. J. D {\bf{7}},  425 (1999).

\bibitem{bogo47}
N. Bogolubov,
\newblock J. Phys. (Moscow) {\bf{11}},  23 (1947).

\bibitem{pita61}
L. P. Pitaevskii,
\newblock J. Exptl. Theoret. Phys. {\bf{40}},  646 (1996) 
[Sov. Phys. JETP {\bf{13}}, 451 (1961)].

\bibitem{gros61}
E. P. Gross,
\newblock Nuovo Cimento {\bf{20}},  454 (1961).

\bibitem{gros63}
E. P. Gross,
\newblock J. Math. Phys. {\bf{4}},  195 (1963).

\bibitem{huan57}
K. Huang and C.~N. Yang,
\newblock Phys. Rev. {\bf{105}},   767 (1957).

\bibitem{lee57}
T.~D. Lee and C.~N. Yang,
\newblock Phys. Rev. {\bf{105}},   1119 (1957).

\bibitem{lee57a}
T.~D. Lee, K. Huang and C.~N. Yang,
\newblock Phys. Rev. {\bf{106}},  1135 (1957).

\bibitem{huan63}
K. Huang,
Statistical mechanics,
\newblock John Wiley and Sons, New York (1963).

\bibitem{fett71}
A. L. Fetter and J. D. Walecka, 
Quantum theory of many-particle systems,
\newblock McGraw-Hill, New York (1971).

\bibitem{ferm34}
E. Fermi,
\newblock Nuovo Cimento {\bf{11}}, 157 (1934).

\bibitem{braa97}
E. Braaten and A. Nieto,
\newblock Phys. Rev. B {\bf{56}}, 14745 (1997).

\bibitem{timm97}
E. Timmermans, P. Tommasini and K. Huang,
\newblock Phys. Rev. A {\bf{55}},  3645 (1997).

\bibitem{ande99}
J.~O. Anderson and E. Braaten,
\newblock Phys. Rev. A {\bf{60}},  2330 (1999).

\bibitem{krau96}
W. Krauth,
\newblock Phys. Rev. Lett. {\bf{77}},  3695 (1996).

\bibitem{grut97}
P. Gr\"uter, D. Ceperley and F. Lalo\"e,
\newblock Phys. Rev. Lett. {\bf{79}},  3549 (1997).

\bibitem{fabr99}
A. Fabrocini and A. Polls,
\newblock Phys. Rev. A {\bf{60}},  2319 (1999).

\bibitem{dubo00}
J.~L. DuBois and H.~R. Glyde,
\newblock cond-mat/0008368 (unpublished).

\bibitem{zalu00}
M.~A. Za{\l}uska-Kotur, M. Gajda, A. Or{\l}owski and J. Mostowski,
\newblock Phys. Rev. A {\bf{61}}, 033613 (2000).

\bibitem{zieg97}
K. Ziegler and A. Shukla,
\newblock Phys. Rev. A {\bf{56}},  1438 (1997).

\bibitem{hutc98}
D. A. W. Hutchinson, R. J. Dodd and K. Burnett,
\newblock Phys. Rev. Lett. {\bf{81}},  2198 (1998).

\bibitem{morg00}
S. A. Morgan,
\newblock J. Phys. B: At. Mol. Opt. Phys. {\bf{33}},  3847 (2000).

\bibitem{hamm94}
B.~L. Hammond, W.~A. {Lester, Jr.} and P.~J. Reynolds,
\newblock Monte Carlo Methods in Ab Initio Quantum Chemistry,
\newblock (World Scientific, Singapore, 1994).

\bibitem{krau90}
M. Krauss and W.~J. Stevens,
\newblock J. Chem. Phys. {\bf{93}},   4236 (1990).

\bibitem{gior99}
S. Giorgini, J. Boronat and J. Casulleras,
\newblock Phys. Rev. A {\bf{60}},  5129 (1999).

\bibitem{pita98}
L. Pitaevskii and S. Stringari,
\newblock Phys. Rev. Lett. {\bf{81}},   4541 (1998).

\bibitem{braa99a}
E. Braaten and J. Pearson,
\newblock Phys. Rev. Lett. {\bf{82}},   255 (1999).

\bibitem{corn00}
S.~L. Cornish, N.~R. Claussen, J.~L. Roberts, E.~A. Cornell and C.~E. Wieman,
\newblock Phys. Rev. Lett. {\bf{85}},  1795 (2000).

\end{thebibliography}
\end{document}